\documentclass[aps,prl,twocolumn,showpacs,superscriptaddress,groupedaddress]{revtex4}  
\usepackage{graphicx}  
\usepackage{dcolumn}   
\usepackage{bm}        
\usepackage{amssymb}   

\usepackage{color}
\definecolor{darkgreen}{rgb}{0,0.7,0}

\newcommand{\ie}{\textit{i.e.,~}}
\newcommand{\eg}{\textit{e.g.,~}}

\newcommand{\nfnl}{n_{f_\mathrm{NL}}}
\newcommand{\fnl}{f_\mathrm{NL}}
\newcommand{\gnl}{g_\mathrm{NL}}

\newcommand{\equ}[1]{\begin{equation}#1\end{equation}}
\newcommand{\eqn}[1]{\begin{eqnarray}#1\end{eqnarray}}

\begin{document}

\title{Constraints on primordial non-Gaussianity from 800,000 photometric quasars}

\author{Boris Leistedt}
  \email{boris.leistedt.11@ucl.ac.uk}
  \affiliation{Department of Physics and Astronomy, University College London, London WC1E 6BT, U.K.}

\author{Hiranya V. Peiris}
  \email{h.peiris@ucl.ac.uk}
  \affiliation{Department of Physics and Astronomy, University College London, London WC1E 6BT, U.K.}

\author{Nina Roth}
  \email{n.roth@ucl.ac.uk}
  \affiliation{Department of Physics and Astronomy, University College London, London WC1E 6BT, U.K.}
  
\date{\today}

\begin{abstract}
We derive robust constraints on primordial non-Gaussianity (PNG) using the clustering of 800,000 photometric quasars from the Sloan Digital Sky Survey in the redshift range $0.5<z<3.5$. These measurements rely on the novel technique of {\it extended mode projection} to control the impact of spatially-varying systematics in a robust fashion, making use of blind analysis techniques. This allows the accurate measurement of quasar halo bias at the largest scales, while discarding as little as possible of the data. The standard local-type PNG parameters $f_\mathrm{NL}$ and $g_\mathrm{NL}$ both imprint a $k^{-2}$ scale-dependent effect in the bias. Constraining these individually, we obtain $-49<f_\mathrm{NL}<31$ and $-2.7\times10^5<g_\mathrm{NL}<1.9\times10^5$, while their joint constraints lead to $-105<f_\mathrm{NL}<72$ and $-4.0\times10^5<g_\mathrm{NL}<4.9\times10^5$ (all at 95\% CL) . Introducing a running parameter $n_{f_\mathrm{NL}}$ to constrain $b(k) \propto k^{-2+n_{f_\mathrm{NL}}}$ and a generalised PNG amplitude $\tilde{f}_\mathrm{NL}$,  we obtain $-45.5 \exp({3.7\, n_{f_\mathrm{NL}}}) < \tilde{f}_\mathrm{NL} < 34.4 \exp({3.3\, n_{f_\mathrm{NL}}})$ at 95\% CL. These results incorporate uncertainties in the cosmological parameters, redshift distributions, shot noise, and the bias prescription used to relate the quasar clustering to the underlying dark matter. These are the strongest constraints obtained to date on PNG using a single population of large-scale structure tracers, and are already at the level of pre-{\it Planck} constraints from the cosmic microwave background. A conservative forecast for a {\it Large Synoptic Survey Telescope}-like survey incorporating mode projection yields $\sigma(\fnl) \sim 5$ -- competitive with the {\it Planck} result -- highlighting the power of upcoming large scale structure surveys to probe the initial conditions of the universe. 
\end{abstract}

\maketitle

Canonical single-field slow-roll inflation predicts initial conditions for structure formation that are essentially Gaussian \cite{1987PhLB..197...66A,1990PhRvD..42.3936S,1993ApJ...403L...1F,GanguiLucchin1994,2000MNRAS.313..141V,2000PhRvD..61f3504W,2000MNRAS.313..323G,2001PhRvD..63f3002K,2003JHEP...05..013M,2003NuPhB.667..119A,2004JCAP...08..009B,2004PhR...402..103B}. Any measurement of deviations from this prediction --- summarised by the term primordial non-Gaussianity (PNG) --- can thus provide evidence for non-standard inflationary physics. One of the most physically interesting forms of PNG is the so-called \emph{local} model, where the primordial potential $\phi$ is modified by including higher order terms, 
\equ{
	\Phi = \phi + \fnl [\phi^2 - \langle \phi^2 \rangle ] + \gnl [\phi^3 - 3\phi \langle\phi^2\rangle], \label{pngdef}
}
where all fields are evaluated at the same spatial coordinate, and $\fnl$ and $\gnl$ are real-valued constants (often called the \emph{skewness} and \emph{kurtosis} parameters). 

The most stringent constraints on PNG currently come from higher-order statistics of the cosmic microwave background (CMB). Most recently, the \emph{Planck} collaboration reported $-8.9 < \fnl < 14.3$ (95\% CL) \cite{Planck2013nongaussianity}, while constraints on the kurtosis have been obtained from the WMAP satellite: $-7.4 \times 10^5 < \gnl < 8.2 \times 10^{5}$ (WMAP5, 95\% CL) \cite{SmidtAmblard2010}, or $\gnl = (-4.3\pm 2.3) \times 10^5$ (WMAP9, 68\% CL) \cite{Regan:2013jua}, $\gnl = (-3.3\pm 2.2) \times 10^5$ (WMAP9, 68\% CL) \cite{SekiguchiSugiyama2013}. While these results are compatible with Gaussian initial conditions, their uncertainties still leave room for non-standard inflation models.

The unknown relation (\emph{bias}) between the dark matter density field and a set of observed tracers (which inhabit dark matter halos) is generally considered to be a \emph{complication} in constraining cosmological parameters from large-scale structure (LSS) data. However, in the case of PNG, the bias is actually an \emph{advantage} that can be used to distinguish between non-Gaussianity in the initial conditions and that generated through late-time non-linear structure formation. PNG introduces a distinctive $k$-dependence into the halo bias; qualitatively, the bias for local-type PNG scales as $b(k)\sim k^{-2}$ \cite{Dalal2008png,matarrese2008,SlosarHirata2008,DesjacquesSeljak2010,Smith:2011ub}. 

This implies that the strongest signal can be expected on large scales (small $k$), accessible to wide-area galaxy surveys. At these scales, the bias can be well-approximated by a multiplicative factor between the dark matter- and galaxy power spectra. LSS clustering constraints on PNG provide an independent validation of the CMB results, and are predicted to improve significantly with on-going and future LSS surveys, eventually surpassing CMB constraints if systematic errors can be controlled \cite{2008ApJ...684L...1C,2010CQGra..27l4011D,2011ApJ...728L..13M,2011PhRvD..84h3509H,Giannantonio2012,2012MNRAS.422...44P,2013PhRvD..88b3534L,2013PhRvD..88h1303M}. Quasars --  the bright nuclei at the centre of the most active galaxies -- are highly-biased tracers of the LSS, spanning large volumes and covering extended redshift ranges: in principle, quasar surveys are ideal for constraining PNG \cite{SlosarHirata2008, Giannantonio2013png}. However, previous analyses have been complicated by the presence of spurious excess power at large scales due to systematics, which mimic the signature of PNG \cite{Xia2009highzisw, Xia2011sdssqsocell, PullenHirata2012, Giannantonio2013png, Leistedt2013excessdr6, LeistedtPeiris:2014:XDQSO_cls}.

In this {\it Letter}, we use a large sample of quasars \cite{bovy2012xdqsoz} from the Sloan Digital Sky Survey (SDSS) \cite{Gunn2006} to constrain PNG. Using a novel technique for blind mitigation of systematics described in Leistedt \& Peiris (2014) \cite{LeistedtPeiris:2014:XDQSO_cls}, we are able to significantly enhance the constraining power of the dataset, resulting in PNG constraints from a single LSS dataset which are competitive with those from the CMB. Our results represent a significant step toward achieving the exquisite control of systematics necessary to exploit future LSS surveys to measure PNG.

{\bf PNG with photometric quasars.} The first LSS constraints on PNG were derived in Ref.~\cite{SlosarHirata2008} using a combination of tracers from early SDSS releases, leading to $\fnl=28^{+23}_{-24}$ (68\% CL). Among these tracers, the photometric quasars --- candidate quasars identified using imaging data only --- have the highest bias and probe the largest volume. Therefore, they had the most constraining power ($\fnl=8^{+26}_{-37}$ at 68\% CL), demonstrating their potential to constrain PNG and complement CMB experiments. However, subsequent analyses \cite{Xia2009highzisw, Xia2011sdssqsocell, Giannantonio2013png} of photometric quasars from the Sixth SDSS Data Release (using the catalogue from Ref.~\cite{Richards2008rqcat}) also revealed systematic effects, such as spatially-varying depth and stellar contamination, which could strongly bias the clustering measurements on the largest scales and jeopardise cosmological inferences if not properly mitigated \cite{PullenHirata2012, Leistedt2013excessdr6}. Refs.~\cite{Giannantonio2013png, Giannantonio2013crosscmblss} obtained $\fnl=5\pm21$ (68\% CL) using a range of LSS probes, discarding the auto-correlation of quasars to avoid the main systematic contamination. Ref.~\cite{hoagarwal2013xdqsoz} used the latest catalogue of SDSS photometric quasars, XDQSOz \citep{Bovy2010xdqso, bovy2012xdqsoz}, to obtain $\fnl=103^{+148}_{-146}$ (68\% CL), and $\fnl=2^{+65}_{-66}$ (68\% CL) when combined with constraints from the clustering of Luminous Red Galaxies, relying however on stringent quality cuts on the survey maps and power spectra to limit the impact of systematics. Therefore, photometric quasars have thus far remained underexploited for PNG, as systematics are handled by removing hard-won data.

{\bf XDQSOz power spectrum measurements.} A different approach was adopted by two of us in Ref.~\cite{LeistedtPeiris:2014:XDQSO_cls}, where the SDSS XDQSOz catalogue was analysed with minimal quality and sky cuts, leading to a sample of $\sim 800,000$ quasars covering $\sim 8300\ {\rm deg}^2$ (compared to $\sim 400,000$--$500,000$ objects used by previous analyses). This base sample was further separated into four redshift bins by selecting objects with photometric redshift estimates $\hat{z}_p$ in top-hat windows $[0.5, 1.35]$, $[1.35, 1.7]$, $[1.7, 2.2]$, $[2.2,3.5]$. A quadratic maximum likelihood \cite{Teg97, BJK98b} method was used to simultaneously estimate the auto- and cross-angular power spectra of the $z$-binned data. 
To mitigate the impact of systematics in these power spectra, Ref.~\cite{LeistedtPeiris:2014:XDQSO_cls} introduced a novel technique, {\it extended mode projection}, relying on the fact that most potential systematics (\eg observing conditions, calibration) were also measured during SDSS observations, and could therefore be mapped onto the sky. 

We constructed a non-linear, data-driven model of systematics, using $\sim$3,700 orthogonal templates obtained by decorrelating $\sim$20,000 maps of potential contaminants, including 200 base templates constructed from SDSS data and products of pairs. These orthogonal templates were cross-correlated with the XQDSOz data, yielding null tests which are used to select the most significant systematics following the principles of blind analysis. These were then marginalised over via {\it mode projection} \citep{Teg97, THS1998future, Wandelt:2003uk, SlosarSeljak2004modeproj, PullenHirata2012, Leistedt2013excessdr6} in the power spectrum estimator, self-consistently enhancing the estimator variance.
This approach allows precision control over systematics, and yields robust measurements of the angular power spectra of XDQSOz quasars, even at the largest angular scales. 

We now turn to the ingredients and models needed to connect the power spectrum measurements to theoretical predictions and constrain PNG.

\begin{figure}
\hspace*{-3mm}\includegraphics[trim = 0.0cm 0.1cm 0.0cm 0cm, clip, width=9cm]{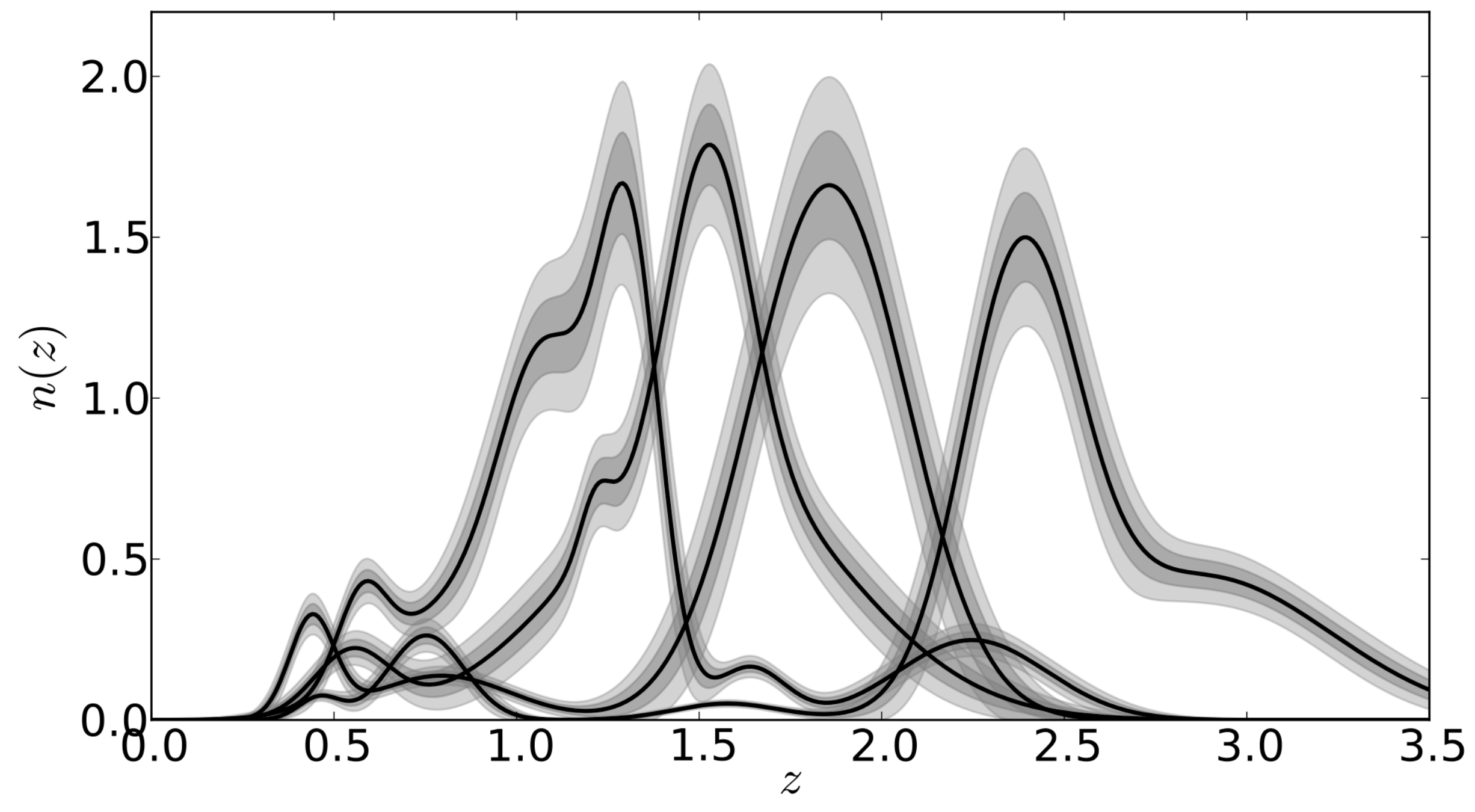}
\caption{Redshift distributions of the four quasar samples used in this analysis, parametrised as superpositions of normal distributions. The shaded regions show the $1,2$ and $3\sigma$ regions explored when adding $5\%$ uncertainty to the parameters of these fits, which are included in the MCMC analysis.}
\label{fig:nz}
\end{figure}

The redshift distributions $n(z)$ of the four quasar samples, shown in Fig.~\ref{fig:nz}, were estimated in Ref.~\cite{LeistedtPeiris:2014:XDQSO_cls} by stacking the posterior distributions of the individual photometric quasars, and then fit with a superposition of Gaussians. Here, we also added a $5\%$ Gaussian uncertainty on the parameters of this fit (illustrated by the shaded bands in Fig.~\ref{fig:nz}) which propagates into the final uncertainties in the PNG parameters.

The shot noise was measured in each sample from the average surface density of photometric quasars, but is also subject to uncertainties due to the unknown fraction of stars in the samples --- between $0$ and $20\%$ from the quality cuts applied to XDQSOz \cite{LeistedtPeiris:2014:XDQSO_cls}. Since non-zero stellar contamination reduces the shot noise, we also marginalised over this effect when constraining PNG.

{\bf Halo bias from PNG.} The impact of local-type PNG is to modify the halo bias by adding a $k$-dependent term to the Gaussian bias $b^G(z)$ \cite{Dalal2008png,matarrese2008,SlosarHirata2008,DesjacquesSeljak2010,Smith:2011ub}
\eqn{
	b^{\rm NG}(k,z) &=& b^{\rm G}(z) + \frac{ \beta_f(z) \fnl + \beta_g(z) \gnl }{\alpha(k,z)}.  \label{equ:pngbias}
}
Here, we neglect an additional small contribution induced by the effect of PNG on the halo mass function, which is independent of $k$ and can thus be absorbed in $b^G(z)$. Note that we have also suppressed the implicit mass dependence of  $b^G$, $\beta_f$, and $\beta_g$ in the previous equation. The exact expression for $\alpha(k,z)$ and fitting functions for $\beta_f$ and $\beta_g$ can be found in Ref.~\cite{Smith:2011ub}.

\begin{figure}
\hspace*{-3mm}\includegraphics[trim = 0.0cm 0cm 0.0cm 0.1cm, clip, width=9cm]{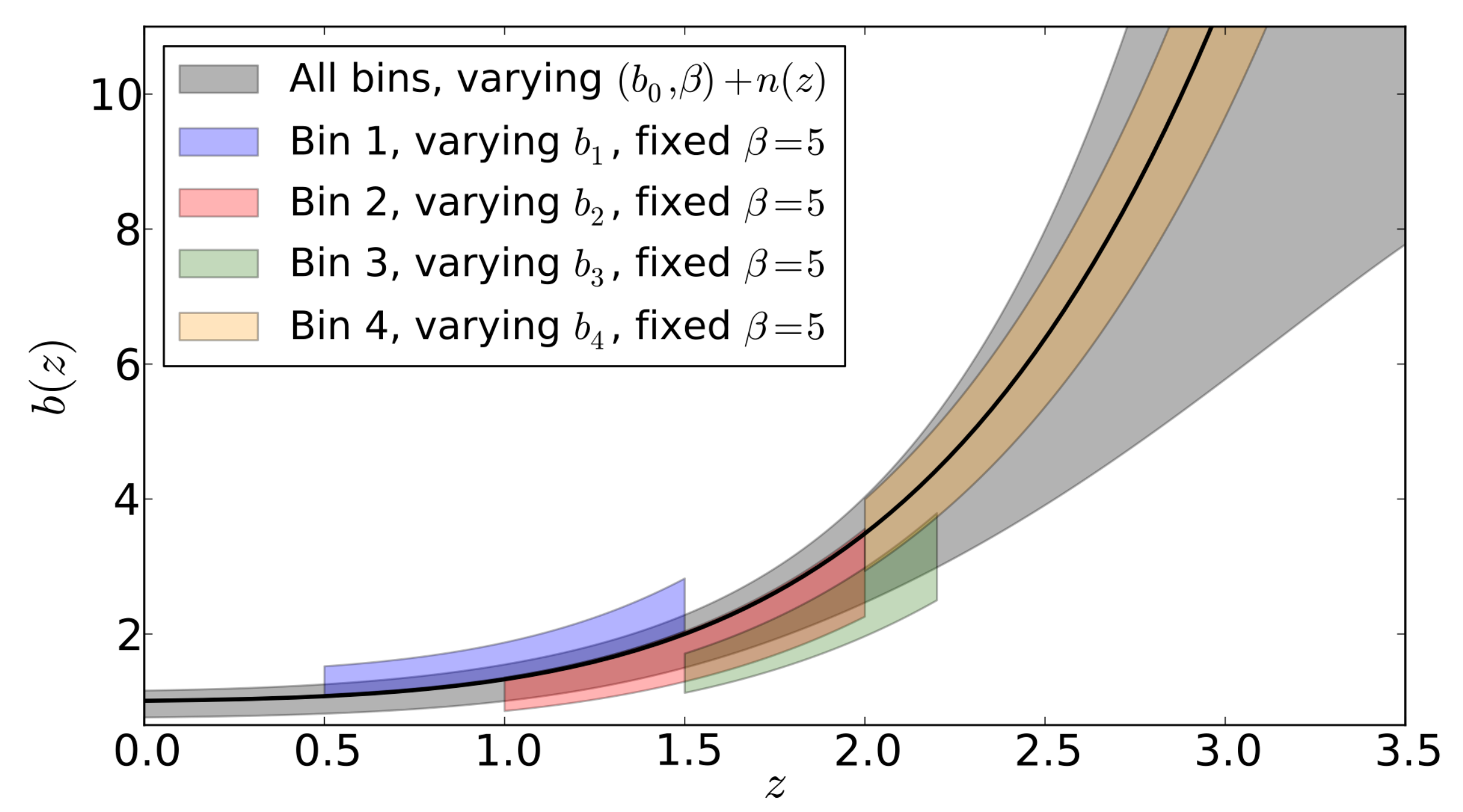}
\caption{Constraints on the quasar bias model described in Eq.~\ref{equ:biasmodel}. The solid line shows the fiducial model with $b_0=1, \beta=5$, and the shaded bands show the $1\sigma$ constraints ($b_0=0.96\pm0.15$, 68\% CL, $\beta$  and $\fnl$ marginalised) from the XDQSOz power spectra when varying the bias and PNG parameters, the shot noise, and the redshift distributions. The coloured bands show the results when fixing $\beta=5$ and allowing a different bias amplitude in each redshift bin, to demonstrate the ability of the overall model to simultaneously describe the four samples.}
\label{fig:biasredshift}
\end{figure}

A simple extension of the local model is the introduction of a spectral index $\nfnl$ in the $\fnl$-generated scale-dependent bias \cite{WagnerVerde2012, 2012JCAP...09..007A, 2013PhRvD..87j7301D, Giannantonio2013png, agarwalho2013xdqsoz}, \ie changing its scaling from $k^{-2}$ into $k^{-2+\nfnl}$ by using
\equ{
	\alpha(k,z) \ \ \ \rightarrow \ \ \ \alpha(k,z) \left( \frac{k}{k_{\rm piv}} \right)^{-\nfnl}, \label{equ:extension}
}
where we choose $k_{\rm piv} = 0.06$ Mpc$^{-1}$. Note that this parametrisation is not equivalent to an intrinsically scale-dependent $\fnl$ as described in Refs.~\cite{Becker:2010hx, Becker:2012yr}. Instead, it allows us to extend our analysis to other types of PNG, like that generated by single-field inflation with a modified initial state \cite{2012JCAP...09..007A}, or models with several light fields \cite{2013PhRvD..87j7301D}.  

The quasar bias is known to evolve strongly with redshift (\eg Refs.~\cite{Myers2006first, Myers2007one, Myers2007two, 2007AJ....133.2222S, White2012specqso}), and thus one cannot use a constant linear bias per redshift bin due to the extended and complicated redshift distributions shown in Fig.~\ref{fig:nz}. For the Gaussian bias $b(z)$ in Eq.~\ref{equ:pngbias}, we used
\equ{
	b(z) = b_0 \left[ 1 + \left(\frac{1+z}{2.5}\right)^{\beta} \right] , \label{equ:biasmodel}
}
which is in good agreement with previous studies of SDSS quasars (\eg Ref.~\cite{porcianiNorberg2006, 2005MNRAS.356..415C, Ross2009specqsodr5}).

\begin{figure}
\hspace*{-4mm}\includegraphics[trim = 0.0cm 0.4cm 0.0cm 0.3cm, clip, width=9.2cm]{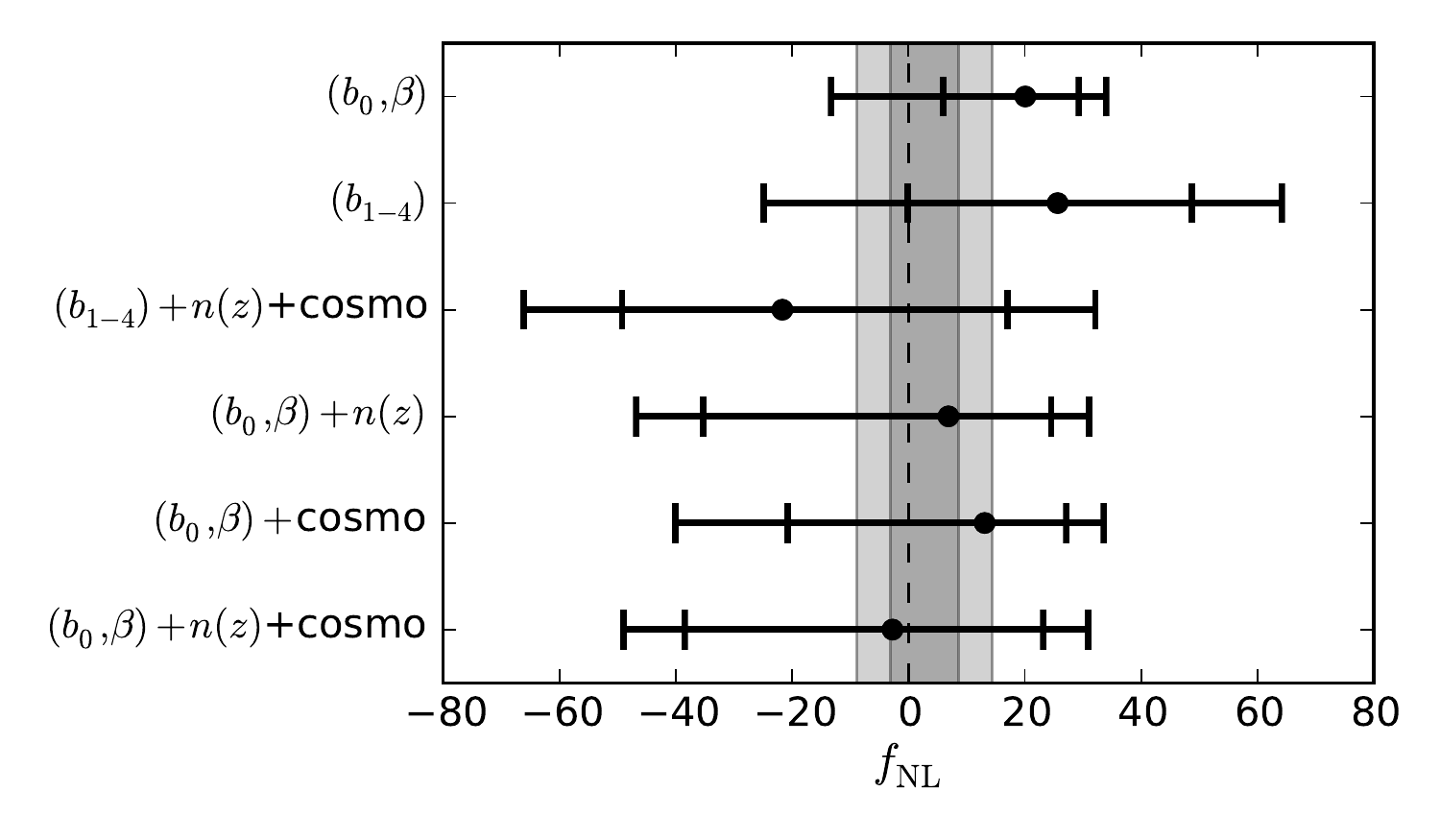}
\caption{Constraints on local-type $\fnl$ (in the $\Lambda$CDM+$\fnl$ model, with $\nfnl=\gnl=0$) using the power spectrum analysis of XDQSOz quasars, for different bias models and incorporating uncertainties in the redshift distributions and cosmological parameters. The error bars show the 1 and 2$\sigma$ constraints, the dashed line shows $\fnl=0$, and the shaded bands show the constraints from {\it Planck} \cite{Planck2013nongaussianity}.}
\label{fig:fnlsummary}
\end{figure}

{\bf Monte Carlo Markov Chain (MCMC) analysis}. We built a Gaussian likelihood \cite{Leistedt2013excessdr6}, jointly using the 10 auto- and cross-angular power spectra (between redshift bins) estimated in Ref.~\cite{LeistedtPeiris:2014:XDQSO_cls}, at multipole resolution $\Delta\ell=15$. The theoretical predictions were calculated using {\tt CAMB\_sources} \citep{challinorlewis2011cambsources}, modified to support PNG and our quasar bias model. We used {\tt emcee} \citep{ForemanMackey2013emcee} to run an MCMC analysis, and sample combinations of the following parameters: 
\underline{Cosmological parameters} (`cosmo'): parameters of the base $\Lambda$CDM model, with fiducial values and uncertainties corresponding to the constraints from {\it Planck} combined with Baryon Acoustic Oscillations (BAO), as in Ref.~\citep{Planck2013cosmologicalparams}. 
\underline{Bias model}: the model described above, with uniform priors $b_0\in[0,2]$ and $\beta\in[4,6]$. 
\underline{Redshift distributions} (`n(z)'): the amplitude and width of the Gaussian functions used to fit the $n(z)$ estimates, with Gaussian priors of $5\%$ $1\sigma$ uncertainties around the fiducial values. Additionally, we sampled the slope of number counts, which controls magnification bias, with Gaussian priors centred at the measured value with $5\%$ $1\sigma$ uncertainty.
\underline{Shot noise}: we marginalised over the shot noise with a prior $[0.8, 1.0]$ times the value measured from the photometric quasar surface density, in order to account for the unknown (but bounded) amount of stellar contamination. 

{\bf Results.} 
We first test the robustness of the bias model by examining the bias measured in the four redshift samples individually and jointly. Therefore, in addition to the `coupled' model presented above, used to connect all power spectra to the theory predictions, we consider an alternative, `decoupled' case where the bias amplitude of each redshift sample is fit separately, using four parameters $b_1, b_2, b_3, b_4$. In this case, we used $\beta=5$ and uniform priors $b_i\in[0,2]$. 
The constraints on the bias parameters from the XDQSOz power spectra are shown in Fig.~\ref{fig:biasredshift}, and demonstrate that the separate bias amplitudes $b_i, i=1\dots4$ of the four samples are in good agreement with each other, with the fiducial model with $b_0=1$ and $\beta=5$ (black line), and also with the results obtained with the coupled model (shaded band). Note that the slope parameter $\beta$ is used to capture the uncertainty in the evolution of the bias at $z>2.5$. This redshift range is not as strongly constrained by the data, but is nevertheless crucial since it has the greatest bias, and therefore is expected to produce the strongest PNG signature.

Having confirmed that Eq.~\ref{equ:biasmodel} is an adequate model for the quasar bias, we can now advance towards constraining the PNG parameters. Unless stated otherwise, all values are quoted at $95\%$ CL.

\begin{figure}
\hspace*{-3mm}\includegraphics[trim = 0.5cm 0.5cm 0.4cm 0.4cm, clip, width=8cm]{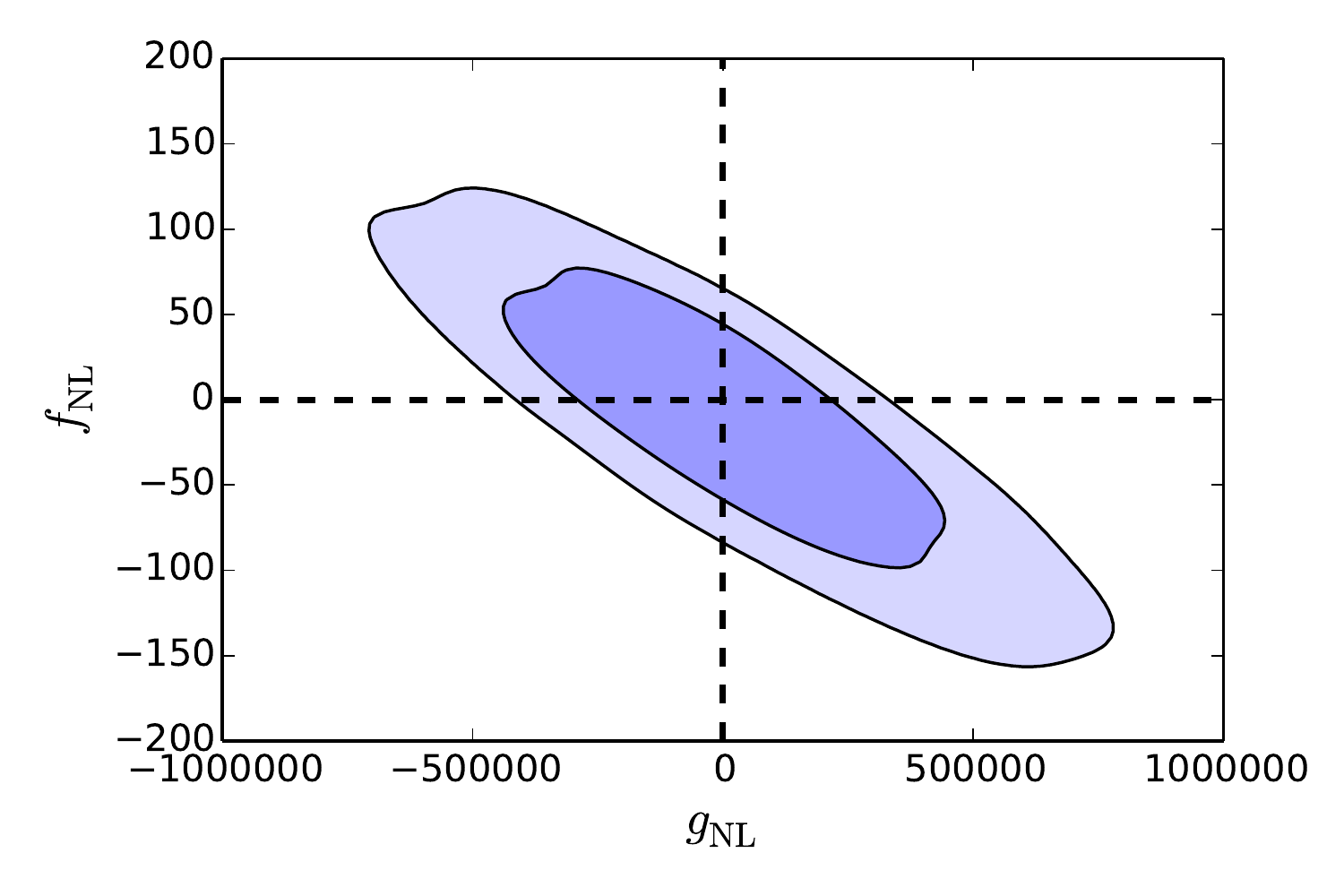}
\caption{$1\sigma$ and $2\sigma$ joint constraints on $\fnl$ and $\gnl$ for the $(b_0,\beta)+n(z)$+cosmo case, \ie marginalising over the uncertainties in the cosmological parameters, redshift distributions, and bias model. }
\label{fig:fnlgnl}
\end{figure}
\begin{figure}
\hspace*{-3mm}\includegraphics[trim = 0.4cm 0.5cm -0.2cm 0.4cm, clip, width=8cm]{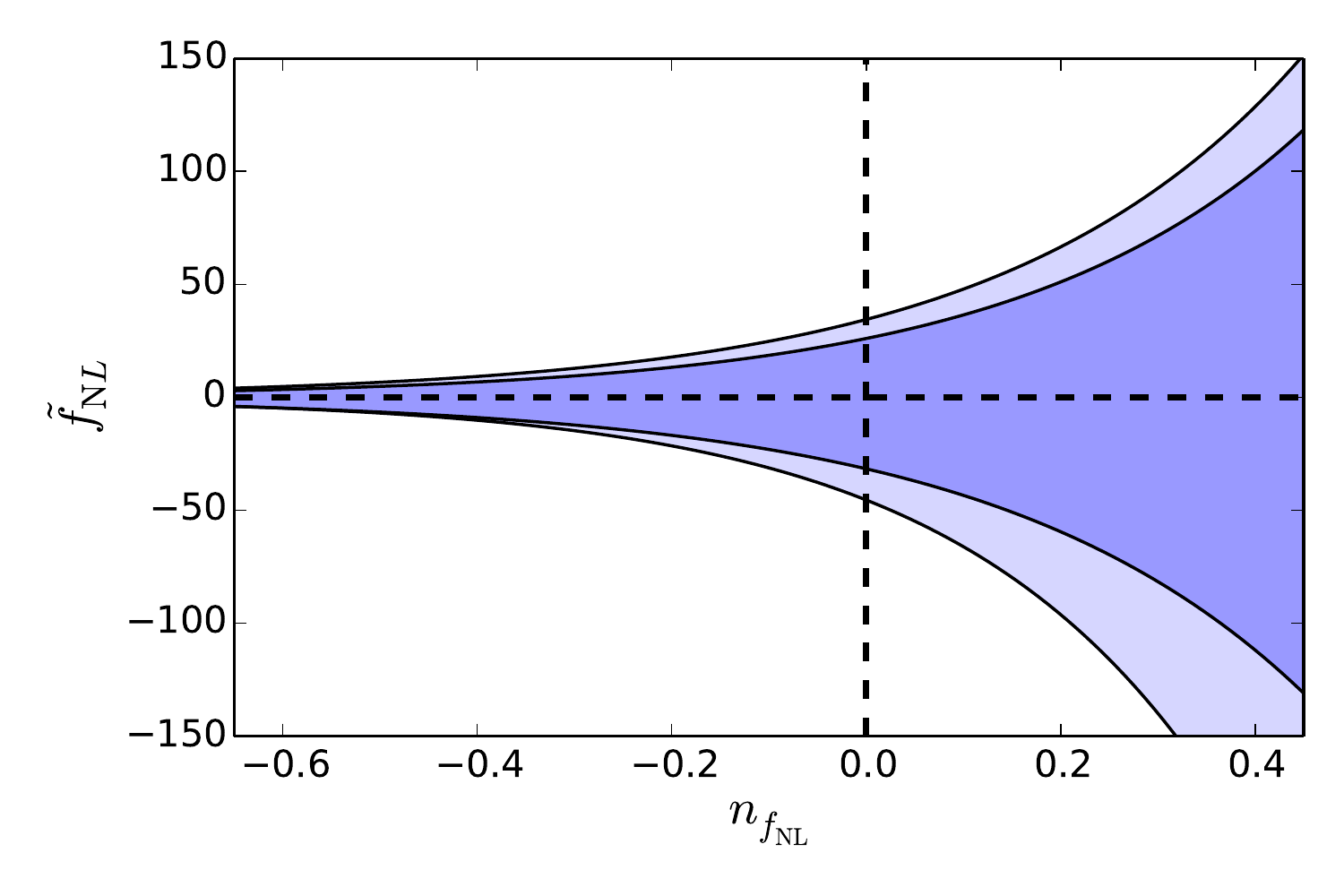}
\caption{$1\sigma$ and $2\sigma$ joint constraints on $\tilde{f}_{\rm{NL}}$ and $\nfnl$ for the extended model of Eq.~\ref{equ:extension}, marginalising over the same parameters as Fig.~\ref{fig:fnlgnl}.  }
\label{fig:fnlns}
\end{figure}

Fig.~\ref{fig:fnlsummary} shows the constraints on $\fnl$ only (with $\nfnl=\gnl=0$) for various combinations of parameters and sources of uncertainties. As expected, adding parameters (\eg using the `decoupled' bias model) increases the error estimates. 
When only varying $b_0, \beta$, we obtain $-26<\fnl<34$ (top row), and adding the uncertainties in the cosmological parameters and redshift distributions relaxes the constraints to $-49<\fnl<31$ (bottom row). Note that this case is more conservative but introduces degeneracies between parameters, in particular those affecting the overall amplitude of the power spectra, such as the bias $b_0$, the amplitude of matter fluctuations $\sigma_8$, and the relative amplitudes of the peaks in $n(z)$. However, $\fnl$ does not suffer from any direct degeneracies with these parameters given its specific signature on large scales, yielding constraints which are robust to the choice of model.

\begin{table}\def\arraystretch{1.5}
\begin{tabular}{| l | c | c | }
\hline
\quad\quad Model  & $\fnl$ & $\gnl / 10^5$  \\\hline 
base+$\fnl$ 	&	$-4_{-35-10}^{+27+8}$ & ---  \\ 
base+$\gnl$  	&	--- &	$0.3_{-2.2-0.8}^{+1.1+0.5}$	\\ 
base+$\fnl$+$\gnl$	&	$-10_{-60-35}^{+54+28}$	&	$0.2_{-2.7-1.5}^{+2.9+1.8}$	 \\ \hline
\end{tabular}
\caption{PNG constraints for different models. The base model is $(b_0,\beta)$+cosmo+$n(z)$. The first super/subscript corresponds to the 68\% CL, and the 95\% CL can be obtained by adding the second number. }
\label{tab:pngvalues}
\end{table}

When only using $\gnl$ as a source of scale-dependent bias, we obtain $-2.7\times10^5<\gnl<1.9\times10^5$. However, as apparent from Eq.~\ref{equ:pngbias}, and as shown in Ref.~\cite{Roth:2012yy}, $\fnl$ and $\gnl$ leave similar signatures on the scale-dependent bias of LSS tracers, and are not easily separable. Despite the large volume and extended redshift range spanned by our data, this degeneracy is confirmed by their joint constraints, shown in Fig.~\ref{fig:fnlgnl}, leading to $-105<\fnl<72$ and $-4.0\times10^5<\gnl<4.9\times10^5$.

For the extended model of Eq.~\ref{equ:extension}, we can introduce a generalised parameter $\tilde{f}_{\rm{NL}}$, which only corresponds to $\fnl$ as defined in Eq.~\ref{pngdef} when $\nfnl=0$. 
In this model, the scale-dependent bias scales as $b(k)\sim k^{-2+\nfnl}$, so the constraints on $\tilde{f}_{\rm{NL}}$ widen as $\nfnl$ increases, due to the less pronounced signature of PNG on large scales. Nevertheless, $\tilde{f}_{\rm{NL}}$ can be constrained at any $\nfnl<2$. Fig.~\ref{fig:fnlns} shows the joint constraints on $\tilde{f}_{\rm{NL}}$ and $\nfnl$, summarised as $-45.5 \exp({3.7\, n_{f_\mathrm{NL}}}) < \tilde{f}_\mathrm{NL} < 34.4 \exp({3.3\, n_{f_\mathrm{NL}}})$, fully compatible with $\tilde{f}_{\rm{NL}}=0$.

\textbf{Future prospects.}
We performed a Fisher forecast of a {\it Large Synoptic Survey Telescope}-like photometric survey \cite{2012arXiv1211.0310L}, using angular power spectra of galaxies in 20 tomographic bins in the redshift range $0.5<z<3.5$, with bias modelled by $b_0 = 1$, $\beta = 3$ in Eq.~\ref{equ:biasmodel}. This idealised forecast, without any contamination by systematics, yields a $1\sigma $ uncertainty on $\fnl$ of $\sim 1$, consistent with the 3D power spectrum analysis of Ref.~\cite{2008ApJ...684L...1C}. 

We then incorporated realistic estimates of the main 50 systematics identified in Ref.~\cite{LeistedtPeiris:2014:XDQSO_cls}. This translates into a $\sim 10\%$ clustering contamination for $\ell >100$, up to factors of a few on the largest scales where the PNG signal is strongest. Without any mitigation in the subsequent analysis, the resulting estimate for $\fnl$ is biased by $\Delta \fnl \sim 30$, which is highly significant compared to the expected uncertainty. Including a harmonic-space mode projection in the forecast removes this bias, while consistently increasing the parameter uncertainty to $\sigma(\fnl)=5$, comparable to current {\it Planck} constraints. However, this increase in uncertainty will be significantly reduced when performing the mode projection in pixel space, using sky maps of the systematics which will be available for a real survey \cite{Leistedt2013excessdr6,LeistedtPeiris:2014:XDQSO_cls}.

\textbf{Conclusion.} Photometric quasar surveys, while potentially constituting ideal datasets for probing PNG, have thus far been systematics-limited. We derived constraints on PNG using the results of a novel power-spectrum estimation method applied to $\sim 800,000$ photometric quasars from SDSS. This approach self-consistently marginalises over a non-linear data-driven model of spatially-varying systematics. Our results, summarised in Table~\ref{tab:pngvalues}, incorporate uncertainties in the cosmological parameters and the parameters of a realistic bias model, while marginalising over uncertainties in the redshift distributions of the quasars. The resulting constraints on $\fnl$ and $\gnl$ are the tightest obtained using a single population of LSS tracers, and are at the level of pre-{\it Planck} CMB constraints. Our results demonstrate the potential of future LSS surveys to reach the $\fnl \sim 1$ levels predicted by the simplest models of inflation.

\mbox{}

\acknowledgments

\paragraph{Acknowledgments.---} 
We are grateful to Jo Bovy, David Hogg and Adam Myers for sharing the XDQSOz catalogue with us, and Jo Bovy, Masahiro Takada for useful conversations.  

BL is supported by the Perren Fund and the IMPACT Fund. HVP and NR are supported by STFC and the European Research Council under the European Community's Seventh Framework Programme (FP7/2007- 2013) / ERC grant agreement no 306478-CosmicDawn. 

We acknowledge use of the following public software packages: \textsc{healpix} \citep{healpix1} and \textsc{camb\_sources} \citep{challinorlewis2011cambsources}. We acknowledge use of the Legacy Archive for Microwave Background Data Analysis (LAMBDA). Support for LAMBDA is provided by the NASA Office of Space Science.  This work is based on observations obtained with SDSS. Funding for the SDSS and SDSS-II has been provided by the Alfred P. Sloan Foundation, the Participating Institutions, the National Science Foundation, the U.S. Department of Energy, the National Aeronautics and Space Administration, the Japanese Monbukagakusho, the Max Planck Society, and the Higher Education Funding Council for England. The SDSS Web Site is http://www.sdss.org/.

\bibliography{bib}

\begin{thebibliography}{66}
\expandafter\ifx\csname natexlab\endcsname\relax\def\natexlab#1{#1}\fi
\expandafter\ifx\csname bibnamefont\endcsname\relax
  \def\bibnamefont#1{#1}\fi
\expandafter\ifx\csname bibfnamefont\endcsname\relax
  \def\bibfnamefont#1{#1}\fi
\expandafter\ifx\csname citenamefont\endcsname\relax
  \def\citenamefont#1{#1}\fi
\expandafter\ifx\csname url\endcsname\relax
  \def\url#1{\texttt{#1}}\fi
\expandafter\ifx\csname urlprefix\endcsname\relax\def\urlprefix{URL }\fi
\providecommand{\bibinfo}[2]{#2}
\providecommand{\eprint}[2][]{\url{#2}}

\bibitem[{\citenamefont{{Allen} et~al.}(1987)\citenamefont{{Allen},
  {Grinstein}, and {Wise}}}]{1987PhLB..197...66A}
\bibinfo{author}{\bibfnamefont{T.~J.} \bibnamefont{{Allen}}},
  \bibinfo{author}{\bibfnamefont{B.}~\bibnamefont{{Grinstein}}},
  \bibnamefont{and} \bibinfo{author}{\bibfnamefont{M.~B.}
  \bibnamefont{{Wise}}}, \bibinfo{journal}{Physics Letters B}
  \textbf{\bibinfo{volume}{197}}, \bibinfo{pages}{66} (\bibinfo{year}{1987}).

\bibitem[{\citenamefont{{Salopek} and {Bond}}(1990)}]{1990PhRvD..42.3936S}
\bibinfo{author}{\bibfnamefont{D.~S.} \bibnamefont{{Salopek}}}
  \bibnamefont{and} \bibinfo{author}{\bibfnamefont{J.~R.}
  \bibnamefont{{Bond}}}, \bibinfo{journal}{\prd} \textbf{\bibinfo{volume}{42}},
  \bibinfo{pages}{3936} (\bibinfo{year}{1990}).

\bibitem[{\citenamefont{{Falk} et~al.}(1993)\citenamefont{{Falk}, {Rangarajan},
  and {Srednicki}}}]{1993ApJ...403L...1F}
\bibinfo{author}{\bibfnamefont{T.}~\bibnamefont{{Falk}}},
  \bibinfo{author}{\bibfnamefont{R.}~\bibnamefont{{Rangarajan}}},
  \bibnamefont{and}
  \bibinfo{author}{\bibfnamefont{M.}~\bibnamefont{{Srednicki}}},
  \bibinfo{journal}{\apjl} \textbf{\bibinfo{volume}{403}}, \bibinfo{pages}{L1}
  (\bibinfo{year}{1993}), \eprint{astro-ph/9208001}.

\bibitem[{\citenamefont{{Gangui} et~al.}(1994)\citenamefont{{Gangui},
  {Lucchin}, {Matarrese}, and {Mollerach}}}]{GanguiLucchin1994}
\bibinfo{author}{\bibfnamefont{A.}~\bibnamefont{{Gangui}}},
  \bibinfo{author}{\bibfnamefont{F.}~\bibnamefont{{Lucchin}}},
  \bibinfo{author}{\bibfnamefont{S.}~\bibnamefont{{Matarrese}}},
  \bibnamefont{and}
  \bibinfo{author}{\bibfnamefont{S.}~\bibnamefont{{Mollerach}}},
  \bibinfo{journal}{\apj} \textbf{\bibinfo{volume}{430}}, \bibinfo{pages}{447}
  (\bibinfo{year}{1994}), \eprint{astro-ph/9312033}.

\bibitem[{\citenamefont{{Verde} et~al.}(2000)\citenamefont{{Verde}, {Wang},
  {Heavens}, and {Kamionkowski}}}]{2000MNRAS.313..141V}
\bibinfo{author}{\bibfnamefont{L.}~\bibnamefont{{Verde}}},
  \bibinfo{author}{\bibfnamefont{L.}~\bibnamefont{{Wang}}},
  \bibinfo{author}{\bibfnamefont{A.~F.} \bibnamefont{{Heavens}}},
  \bibnamefont{and}
  \bibinfo{author}{\bibfnamefont{M.}~\bibnamefont{{Kamionkowski}}},
  \bibinfo{journal}{\mnras} \textbf{\bibinfo{volume}{313}},
  \bibinfo{pages}{141} (\bibinfo{year}{2000}), \eprint{astro-ph/9906301}.

\bibitem[{\citenamefont{{Wang} and {Kamionkowski}}(2000)}]{2000PhRvD..61f3504W}
\bibinfo{author}{\bibfnamefont{L.}~\bibnamefont{{Wang}}} \bibnamefont{and}
  \bibinfo{author}{\bibfnamefont{M.}~\bibnamefont{{Kamionkowski}}},
  \bibinfo{journal}{\prd} \textbf{\bibinfo{volume}{61}}, \bibinfo{eid}{063504}
  (\bibinfo{year}{2000}), \eprint{astro-ph/9907431}.

\bibitem[{\citenamefont{{Gangui} and {Martin}}(2000)}]{2000MNRAS.313..323G}
\bibinfo{author}{\bibfnamefont{A.}~\bibnamefont{{Gangui}}} \bibnamefont{and}
  \bibinfo{author}{\bibfnamefont{J.}~\bibnamefont{{Martin}}},
  \bibinfo{journal}{\mnras} \textbf{\bibinfo{volume}{313}},
  \bibinfo{pages}{323} (\bibinfo{year}{2000}), \eprint{astro-ph/9908009}.

\bibitem[{\citenamefont{{Komatsu} and {Spergel}}(2001)}]{2001PhRvD..63f3002K}
\bibinfo{author}{\bibfnamefont{E.}~\bibnamefont{{Komatsu}}} \bibnamefont{and}
  \bibinfo{author}{\bibfnamefont{D.~N.} \bibnamefont{{Spergel}}},
  \bibinfo{journal}{\prd} \textbf{\bibinfo{volume}{63}}, \bibinfo{eid}{063002}
  (\bibinfo{year}{2001}), \eprint{astro-ph/0005036}.

\bibitem[{\citenamefont{{Maldacena}}(2003)}]{2003JHEP...05..013M}
\bibinfo{author}{\bibfnamefont{J.}~\bibnamefont{{Maldacena}}},
  \bibinfo{journal}{Journal of High Energy Physics}
  \textbf{\bibinfo{volume}{5}}, \bibinfo{eid}{013} (\bibinfo{year}{2003}),
  \eprint{astro-ph/0210603}.

\bibitem[{\citenamefont{{Acquaviva} et~al.}(2003)\citenamefont{{Acquaviva},
  {Bartolo}, {Matarrese}, and {Riotto}}}]{2003NuPhB.667..119A}
\bibinfo{author}{\bibfnamefont{V.}~\bibnamefont{{Acquaviva}}},
  \bibinfo{author}{\bibfnamefont{N.}~\bibnamefont{{Bartolo}}},
  \bibinfo{author}{\bibfnamefont{S.}~\bibnamefont{{Matarrese}}},
  \bibnamefont{and} \bibinfo{author}{\bibfnamefont{A.}~\bibnamefont{{Riotto}}},
  \bibinfo{journal}{Nuclear Physics B} \textbf{\bibinfo{volume}{667}},
  \bibinfo{pages}{119} (\bibinfo{year}{2003}), \eprint{astro-ph/0209156}.

\bibitem[{\citenamefont{{Babich} et~al.}(2004)\citenamefont{{Babich},
  {Creminelli}, and {Zaldarriaga}}}]{2004JCAP...08..009B}
\bibinfo{author}{\bibfnamefont{D.}~\bibnamefont{{Babich}}},
  \bibinfo{author}{\bibfnamefont{P.}~\bibnamefont{{Creminelli}}},
  \bibnamefont{and}
  \bibinfo{author}{\bibfnamefont{M.}~\bibnamefont{{Zaldarriaga}}},
  \bibinfo{journal}{\jcap} \textbf{\bibinfo{volume}{8}}, \bibinfo{eid}{009}
  (\bibinfo{year}{2004}), \eprint{astro-ph/0405356}.

\bibitem[{\citenamefont{{Bartolo} et~al.}(2004)\citenamefont{{Bartolo},
  {Komatsu}, {Matarrese}, and {Riotto}}}]{2004PhR...402..103B}
\bibinfo{author}{\bibfnamefont{N.}~\bibnamefont{{Bartolo}}},
  \bibinfo{author}{\bibfnamefont{E.}~\bibnamefont{{Komatsu}}},
  \bibinfo{author}{\bibfnamefont{S.}~\bibnamefont{{Matarrese}}},
  \bibnamefont{and} \bibinfo{author}{\bibfnamefont{A.}~\bibnamefont{{Riotto}}},
  \bibinfo{journal}{\physrep} \textbf{\bibinfo{volume}{402}},
  \bibinfo{pages}{103} (\bibinfo{year}{2004}), \eprint{astro-ph/0406398}.

\bibitem[{\citenamefont{{Planck
  Collaboration}}(2013{\natexlab{a}})}]{Planck2013nongaussianity}
\bibinfo{author}{\bibnamefont{{Planck Collaboration}}}, \bibinfo{journal}{ArXiv
  e-prints}  (\bibinfo{year}{2013}{\natexlab{a}}), \eprint{1303.5084}.

\bibitem[{\citenamefont{Smidt et~al.}(2010)\citenamefont{Smidt, Amblard,
  Byrnes, Cooray, Heavens, and Munshi}}]{SmidtAmblard2010}
\bibinfo{author}{\bibfnamefont{J.}~\bibnamefont{Smidt}},
  \bibinfo{author}{\bibfnamefont{A.}~\bibnamefont{Amblard}},
  \bibinfo{author}{\bibfnamefont{C.~T.} \bibnamefont{Byrnes}},
  \bibinfo{author}{\bibfnamefont{A.}~\bibnamefont{Cooray}},
  \bibinfo{author}{\bibfnamefont{A.}~\bibnamefont{Heavens}}, \bibnamefont{and}
  \bibinfo{author}{\bibfnamefont{D.}~\bibnamefont{Munshi}},
  \bibinfo{journal}{Phys. Rev. D} \textbf{\bibinfo{volume}{81}},
  \bibinfo{pages}{123007} (\bibinfo{year}{2010}).

\bibitem[{\citenamefont{Regan et~al.}(2013)\citenamefont{Regan, Gosenca, and
  Seery}}]{Regan:2013jua}
\bibinfo{author}{\bibfnamefont{D.}~\bibnamefont{Regan}},
  \bibinfo{author}{\bibfnamefont{M.}~\bibnamefont{Gosenca}}, \bibnamefont{and}
  \bibinfo{author}{\bibfnamefont{D.}~\bibnamefont{Seery}}
  (\bibinfo{year}{2013}), \eprint{1310.8617}.

\bibitem[{\citenamefont{Sekiguchi and Sugiyama}(2013)}]{SekiguchiSugiyama2013}
\bibinfo{author}{\bibfnamefont{T.}~\bibnamefont{Sekiguchi}} \bibnamefont{and}
  \bibinfo{author}{\bibfnamefont{N.}~\bibnamefont{Sugiyama}},
  \bibinfo{journal}{Journal of Cosmology and Astroparticle Physics}
  \textbf{\bibinfo{volume}{2013}}, \bibinfo{pages}{002} (\bibinfo{year}{2013}).

\bibitem[{\citenamefont{{Dalal} et~al.}(2008)\citenamefont{{Dalal}, {Dor{\'e}},
  {Huterer}, and {Shirokov}}}]{Dalal2008png}
\bibinfo{author}{\bibfnamefont{N.}~\bibnamefont{{Dalal}}},
  \bibinfo{author}{\bibfnamefont{O.}~\bibnamefont{{Dor{\'e}}}},
  \bibinfo{author}{\bibfnamefont{D.}~\bibnamefont{{Huterer}}},
  \bibnamefont{and}
  \bibinfo{author}{\bibfnamefont{A.}~\bibnamefont{{Shirokov}}},
  \bibinfo{journal}{Phys.\ Rev.\ D.} \textbf{\bibinfo{volume}{77}},
  \bibinfo{eid}{123514} (\bibinfo{year}{2008}), \eprint{0710.4560}.

\bibitem[{\citenamefont{{Matarrese} and {Verde}}(2008)}]{matarrese2008}
\bibinfo{author}{\bibfnamefont{S.}~\bibnamefont{{Matarrese}}} \bibnamefont{and}
  \bibinfo{author}{\bibfnamefont{L.}~\bibnamefont{{Verde}}},
  \bibinfo{journal}{Astrophys.\ J.\ Lett.} \textbf{\bibinfo{volume}{677}},
  \bibinfo{pages}{L77} (\bibinfo{year}{2008}), \eprint{0801.4826}.

\bibitem[{\citenamefont{{Slosar} et~al.}(2008)\citenamefont{{Slosar}, {Hirata},
  {Seljak}, {Ho}, and {Padmanabhan}}}]{SlosarHirata2008}
\bibinfo{author}{\bibfnamefont{A.}~\bibnamefont{{Slosar}}},
  \bibinfo{author}{\bibfnamefont{C.}~\bibnamefont{{Hirata}}},
  \bibinfo{author}{\bibfnamefont{U.}~\bibnamefont{{Seljak}}},
  \bibinfo{author}{\bibfnamefont{S.}~\bibnamefont{{Ho}}}, \bibnamefont{and}
  \bibinfo{author}{\bibfnamefont{N.}~\bibnamefont{{Padmanabhan}}},
  \bibinfo{journal}{Journal of Cosmology and Astroparticle Physics}
  \textbf{\bibinfo{volume}{8}}, \bibinfo{eid}{031} (\bibinfo{year}{2008}),
  \eprint{0805.3580}.

\bibitem[{\citenamefont{{Desjacques} and
  {Seljak}}(2010{\natexlab{a}})}]{DesjacquesSeljak2010}
\bibinfo{author}{\bibfnamefont{V.}~\bibnamefont{{Desjacques}}}
  \bibnamefont{and} \bibinfo{author}{\bibfnamefont{U.}~\bibnamefont{{Seljak}}},
  \bibinfo{journal}{\prd} \textbf{\bibinfo{volume}{81}}, \bibinfo{eid}{023006}
  (\bibinfo{year}{2010}{\natexlab{a}}), \eprint{0907.2257}.

\bibitem[{\citenamefont{Smith et~al.}(2012)\citenamefont{Smith, Ferraro, and
  LoVerde}}]{Smith:2011ub}
\bibinfo{author}{\bibfnamefont{K.~M.} \bibnamefont{Smith}},
  \bibinfo{author}{\bibfnamefont{S.}~\bibnamefont{Ferraro}}, \bibnamefont{and}
  \bibinfo{author}{\bibfnamefont{M.}~\bibnamefont{LoVerde}},
  \bibinfo{journal}{JCAP} \textbf{\bibinfo{volume}{1203}}, \bibinfo{pages}{032}
  (\bibinfo{year}{2012}), \eprint{1106.0503}.

\bibitem[{\citenamefont{{Carbone} et~al.}(2008)\citenamefont{{Carbone},
  {Verde}, and {Matarrese}}}]{2008ApJ...684L...1C}
\bibinfo{author}{\bibfnamefont{C.}~\bibnamefont{{Carbone}}},
  \bibinfo{author}{\bibfnamefont{L.}~\bibnamefont{{Verde}}}, \bibnamefont{and}
  \bibinfo{author}{\bibfnamefont{S.}~\bibnamefont{{Matarrese}}},
  \bibinfo{journal}{\apjl} \textbf{\bibinfo{volume}{684}}, \bibinfo{pages}{L1}
  (\bibinfo{year}{2008}), \eprint{0806.1950}.

\bibitem[{\citenamefont{{Desjacques} and
  {Seljak}}(2010{\natexlab{b}})}]{2010CQGra..27l4011D}
\bibinfo{author}{\bibfnamefont{V.}~\bibnamefont{{Desjacques}}}
  \bibnamefont{and} \bibinfo{author}{\bibfnamefont{U.}~\bibnamefont{{Seljak}}},
  \bibinfo{journal}{Classical and Quantum Gravity}
  \textbf{\bibinfo{volume}{27}}, \bibinfo{eid}{124011}
  (\bibinfo{year}{2010}{\natexlab{b}}), \eprint{1003.5020}.

\bibitem[{\citenamefont{{Marian} et~al.}(2011)\citenamefont{{Marian},
  {Hilbert}, {Smith}, {Schneider}, and {Desjacques}}}]{2011ApJ...728L..13M}
\bibinfo{author}{\bibfnamefont{L.}~\bibnamefont{{Marian}}},
  \bibinfo{author}{\bibfnamefont{S.}~\bibnamefont{{Hilbert}}},
  \bibinfo{author}{\bibfnamefont{R.~E.} \bibnamefont{{Smith}}},
  \bibinfo{author}{\bibfnamefont{P.}~\bibnamefont{{Schneider}}},
  \bibnamefont{and}
  \bibinfo{author}{\bibfnamefont{V.}~\bibnamefont{{Desjacques}}},
  \bibinfo{journal}{\apjl} \textbf{\bibinfo{volume}{728}}, \bibinfo{eid}{L13}
  (\bibinfo{year}{2011}), \eprint{1010.5242}.

\bibitem[{\citenamefont{{Hamaus} et~al.}(2011)\citenamefont{{Hamaus}, {Seljak},
  and {Desjacques}}}]{2011PhRvD..84h3509H}
\bibinfo{author}{\bibfnamefont{N.}~\bibnamefont{{Hamaus}}},
  \bibinfo{author}{\bibfnamefont{U.}~\bibnamefont{{Seljak}}}, \bibnamefont{and}
  \bibinfo{author}{\bibfnamefont{V.}~\bibnamefont{{Desjacques}}},
  \bibinfo{journal}{\prd} \textbf{\bibinfo{volume}{84}}, \bibinfo{eid}{083509}
  (\bibinfo{year}{2011}), \eprint{1104.2321}.

\bibitem[{\citenamefont{{Giannantonio}
  et~al.}(2012)\citenamefont{{Giannantonio}, {Porciani}, {Carron}, {Amara}, and
  {Pillepich}}}]{Giannantonio2012}
\bibinfo{author}{\bibfnamefont{T.}~\bibnamefont{{Giannantonio}}},
  \bibinfo{author}{\bibfnamefont{C.}~\bibnamefont{{Porciani}}},
  \bibinfo{author}{\bibfnamefont{J.}~\bibnamefont{{Carron}}},
  \bibinfo{author}{\bibfnamefont{A.}~\bibnamefont{{Amara}}}, \bibnamefont{and}
  \bibinfo{author}{\bibfnamefont{A.}~\bibnamefont{{Pillepich}}},
  \bibinfo{journal}{\mnras} \textbf{\bibinfo{volume}{422}},
  \bibinfo{pages}{2854} (\bibinfo{year}{2012}), \eprint{1109.0958}.

\bibitem[{\citenamefont{{Pillepich} et~al.}(2012)\citenamefont{{Pillepich},
  {Porciani}, and {Reiprich}}}]{2012MNRAS.422...44P}
\bibinfo{author}{\bibfnamefont{A.}~\bibnamefont{{Pillepich}}},
  \bibinfo{author}{\bibfnamefont{C.}~\bibnamefont{{Porciani}}},
  \bibnamefont{and} \bibinfo{author}{\bibfnamefont{T.~H.}
  \bibnamefont{{Reiprich}}}, \bibinfo{journal}{\mnras}
  \textbf{\bibinfo{volume}{422}}, \bibinfo{pages}{44} (\bibinfo{year}{2012}),
  \eprint{1111.6587}.

\bibitem[{\citenamefont{{Lidz} et~al.}(2013)\citenamefont{{Lidz}, {Baxter},
  {Adshead}, and {Dodelson}}}]{2013PhRvD..88b3534L}
\bibinfo{author}{\bibfnamefont{A.}~\bibnamefont{{Lidz}}},
  \bibinfo{author}{\bibfnamefont{E.~J.} \bibnamefont{{Baxter}}},
  \bibinfo{author}{\bibfnamefont{P.}~\bibnamefont{{Adshead}}},
  \bibnamefont{and}
  \bibinfo{author}{\bibfnamefont{S.}~\bibnamefont{{Dodelson}}},
  \bibinfo{journal}{\prd} \textbf{\bibinfo{volume}{88}}, \bibinfo{eid}{023534}
  (\bibinfo{year}{2013}), \eprint{1304.8049}.

\bibitem[{\citenamefont{{Mao} et~al.}(2013)\citenamefont{{Mao}, {D'Aloisio},
  {Zhang}, and {Shapiro}}}]{2013PhRvD..88h1303M}
\bibinfo{author}{\bibfnamefont{Y.}~\bibnamefont{{Mao}}},
  \bibinfo{author}{\bibfnamefont{A.}~\bibnamefont{{D'Aloisio}}},
  \bibinfo{author}{\bibfnamefont{J.}~\bibnamefont{{Zhang}}}, \bibnamefont{and}
  \bibinfo{author}{\bibfnamefont{P.~R.} \bibnamefont{{Shapiro}}},
  \bibinfo{journal}{\prd} \textbf{\bibinfo{volume}{88}}, \bibinfo{eid}{081303}
  (\bibinfo{year}{2013}), \eprint{1305.0313}.

\bibitem[{\citenamefont{{Giannantonio}
  et~al.}(2013)\citenamefont{{Giannantonio}, {Ross}, {Percival}, {Crittenden},
  {Bacher}, {Kilbinger}, {Nichol}, and {Weller}}}]{Giannantonio2013png}
\bibinfo{author}{\bibfnamefont{T.}~\bibnamefont{{Giannantonio}}},
  \bibinfo{author}{\bibfnamefont{A.~J.} \bibnamefont{{Ross}}},
  \bibinfo{author}{\bibfnamefont{W.~J.} \bibnamefont{{Percival}}},
  \bibinfo{author}{\bibfnamefont{R.}~\bibnamefont{{Crittenden}}},
  \bibinfo{author}{\bibfnamefont{D.}~\bibnamefont{{Bacher}}},
  \bibinfo{author}{\bibfnamefont{M.}~\bibnamefont{{Kilbinger}}},
  \bibinfo{author}{\bibfnamefont{R.}~\bibnamefont{{Nichol}}}, \bibnamefont{and}
  \bibinfo{author}{\bibfnamefont{J.}~\bibnamefont{{Weller}}},
  \bibinfo{journal}{ArXiv e-prints}  (\bibinfo{year}{2013}),
  \eprint{1303.1349}.

\bibitem[{\citenamefont{{Xia} et~al.}(2009)\citenamefont{{Xia}, {Viel},
  {Baccigalupi}, and {Matarrese}}}]{Xia2009highzisw}
\bibinfo{author}{\bibfnamefont{J.-Q.} \bibnamefont{{Xia}}},
  \bibinfo{author}{\bibfnamefont{M.}~\bibnamefont{{Viel}}},
  \bibinfo{author}{\bibfnamefont{C.}~\bibnamefont{{Baccigalupi}}},
  \bibnamefont{and}
  \bibinfo{author}{\bibfnamefont{S.}~\bibnamefont{{Matarrese}}},
  \bibinfo{journal}{Journal of Cosmology and Astroparticle Physics}
  \textbf{\bibinfo{volume}{9}}, \bibinfo{eid}{003} (\bibinfo{year}{2009}),
  \eprint{0907.4753}.

\bibitem[{\citenamefont{{Xia} et~al.}(2011)\citenamefont{{Xia}, {Baccigalupi},
  {Matarrese}, {Verde}, and {Viel}}}]{Xia2011sdssqsocell}
\bibinfo{author}{\bibfnamefont{J.-Q.} \bibnamefont{{Xia}}},
  \bibinfo{author}{\bibfnamefont{C.}~\bibnamefont{{Baccigalupi}}},
  \bibinfo{author}{\bibfnamefont{S.}~\bibnamefont{{Matarrese}}},
  \bibinfo{author}{\bibfnamefont{L.}~\bibnamefont{{Verde}}}, \bibnamefont{and}
  \bibinfo{author}{\bibfnamefont{M.}~\bibnamefont{{Viel}}},
  \bibinfo{journal}{Journal of Cosmology and Astroparticle Physics}
  \textbf{\bibinfo{volume}{8}}, \bibinfo{eid}{033} (\bibinfo{year}{2011}),
  \eprint{1104.5015}.

\bibitem[{\citenamefont{{Pullen} and {Hirata}}(2012)}]{PullenHirata2012}
\bibinfo{author}{\bibfnamefont{A.~R.} \bibnamefont{{Pullen}}} \bibnamefont{and}
  \bibinfo{author}{\bibfnamefont{C.~M.} \bibnamefont{{Hirata}}},
  \bibinfo{journal}{ArXiv e-prints}  (\bibinfo{year}{2012}),
  \eprint{1212.4500}.

\bibitem[{\citenamefont{{Leistedt} et~al.}(2013)\citenamefont{{Leistedt},
  {Peiris}, {Mortlock}, {Benoit-L{\'e}vy}, and
  {Pontzen}}}]{Leistedt2013excessdr6}
\bibinfo{author}{\bibfnamefont{B.}~\bibnamefont{{Leistedt}}},
  \bibinfo{author}{\bibfnamefont{H.~V.} \bibnamefont{{Peiris}}},
  \bibinfo{author}{\bibfnamefont{D.~J.} \bibnamefont{{Mortlock}}},
  \bibinfo{author}{\bibfnamefont{A.}~\bibnamefont{{Benoit-L{\'e}vy}}},
  \bibnamefont{and}
  \bibinfo{author}{\bibfnamefont{A.}~\bibnamefont{{Pontzen}}},
  \bibinfo{journal}{Mon.\ Not.\ Roy.\ Astron.\ Soc.}
  \textbf{\bibinfo{volume}{435}}, \bibinfo{pages}{1857} (\bibinfo{year}{2013}),
  \eprint{1306.0005}.

\bibitem[{\citenamefont{{Leistedt} and
  {Peiris}}(2014)}]{LeistedtPeiris:2014:XDQSO_cls}
\bibinfo{author}{\bibfnamefont{B.}~\bibnamefont{{Leistedt}}} \bibnamefont{and}
  \bibinfo{author}{\bibfnamefont{H.~V.} \bibnamefont{{Peiris}}},
  \bibinfo{journal}{ArXiv e-prints}  (\bibinfo{year}{2014}),
  \eprint{1404.6530}.

\bibitem[{\citenamefont{{Bovy} et~al.}(2012)\citenamefont{{Bovy}, {Myers},
  {Hennawi}, {Hogg}, {McMahon}, {Schiminovich}, {Sheldon}, {Brinkmann},
  {Schneider}, and {Weaver}}}]{bovy2012xdqsoz}
\bibinfo{author}{\bibfnamefont{J.}~\bibnamefont{{Bovy}}},
  \bibinfo{author}{\bibfnamefont{A.~D.} \bibnamefont{{Myers}}},
  \bibinfo{author}{\bibfnamefont{J.~F.} \bibnamefont{{Hennawi}}},
  \bibinfo{author}{\bibfnamefont{D.~W.} \bibnamefont{{Hogg}}},
  \bibinfo{author}{\bibfnamefont{R.~G.} \bibnamefont{{McMahon}}},
  \bibinfo{author}{\bibfnamefont{D.}~\bibnamefont{{Schiminovich}}},
  \bibinfo{author}{\bibfnamefont{E.~S.} \bibnamefont{{Sheldon}}},
  \bibinfo{author}{\bibfnamefont{J.}~\bibnamefont{{Brinkmann}}},
  \bibinfo{author}{\bibfnamefont{D.~P.} \bibnamefont{{Schneider}}},
  \bibnamefont{and} \bibinfo{author}{\bibfnamefont{B.~A.}
  \bibnamefont{{Weaver}}}, \bibinfo{journal}{Astrophys.\ J.}
  \textbf{\bibinfo{volume}{749}}, \bibinfo{eid}{41} (\bibinfo{year}{2012}),
  \eprint{1105.3975}.

\bibitem[{\citenamefont{{Gunn} et~al.}(2006)\citenamefont{{Gunn}, {Siegmund},
  {Mannery}, {Owen}, {Hull}, {Leger}, {Carey}, {Knapp}, {York}, {Boroski}
  et~al.}}]{Gunn2006}
\bibinfo{author}{\bibfnamefont{J.~E.} \bibnamefont{{Gunn}}},
  \bibinfo{author}{\bibfnamefont{W.~A.} \bibnamefont{{Siegmund}}},
  \bibinfo{author}{\bibfnamefont{E.~J.} \bibnamefont{{Mannery}}},
  \bibinfo{author}{\bibfnamefont{R.~E.} \bibnamefont{{Owen}}},
  \bibinfo{author}{\bibfnamefont{C.~L.} \bibnamefont{{Hull}}},
  \bibinfo{author}{\bibfnamefont{R.~F.} \bibnamefont{{Leger}}},
  \bibinfo{author}{\bibfnamefont{L.~N.} \bibnamefont{{Carey}}},
  \bibinfo{author}{\bibfnamefont{G.~R.} \bibnamefont{{Knapp}}},
  \bibinfo{author}{\bibfnamefont{D.~G.} \bibnamefont{{York}}},
  \bibinfo{author}{\bibfnamefont{W.~N.} \bibnamefont{{Boroski}}},
  \bibnamefont{et~al.}, \bibinfo{journal}{Astrophys.\ J.}
  \textbf{\bibinfo{volume}{131}}, \bibinfo{pages}{2332} (\bibinfo{year}{2006}),
  \eprint{astro-ph/0602326}.

\bibitem[{\citenamefont{{Richards} et~al.}(2009)\citenamefont{{Richards},
  {Myers}, {Gray}, {Riegel}, {Nichol}, {Brunner}, {Szalay}, {Schneider}, and
  {Anderson}}}]{Richards2008rqcat}
\bibinfo{author}{\bibfnamefont{G.~T.} \bibnamefont{{Richards}}},
  \bibinfo{author}{\bibfnamefont{A.~D.} \bibnamefont{{Myers}}},
  \bibinfo{author}{\bibfnamefont{A.~G.} \bibnamefont{{Gray}}},
  \bibinfo{author}{\bibfnamefont{R.~N.} \bibnamefont{{Riegel}}},
  \bibinfo{author}{\bibfnamefont{R.~C.} \bibnamefont{{Nichol}}},
  \bibinfo{author}{\bibfnamefont{R.~J.} \bibnamefont{{Brunner}}},
  \bibinfo{author}{\bibfnamefont{A.~S.} \bibnamefont{{Szalay}}},
  \bibinfo{author}{\bibfnamefont{D.~P.} \bibnamefont{{Schneider}}},
  \bibnamefont{and} \bibinfo{author}{\bibfnamefont{S.~F.}
  \bibnamefont{{Anderson}}}, \bibinfo{journal}{Astrophys.\ J.\ Supp.}
  \textbf{\bibinfo{volume}{180}}, \bibinfo{pages}{67} (\bibinfo{year}{2009}),
  \eprint{0809.3952}.

\bibitem[{\citenamefont{{Giannantonio} and
  {Percival}}(2013)}]{Giannantonio2013crosscmblss}
\bibinfo{author}{\bibfnamefont{T.}~\bibnamefont{{Giannantonio}}}
  \bibnamefont{and} \bibinfo{author}{\bibfnamefont{W.~J.}
  \bibnamefont{{Percival}}}, \bibinfo{journal}{ArXiv e-prints}
  (\bibinfo{year}{2013}), \eprint{1312.5154}.

\bibitem[{\citenamefont{{Ho} et~al.}(2013)\citenamefont{{Ho}, {Agarwal},
  {Myers}, {Lyons}, {Disbrow}, {Seo}, {Ross}, {Hirata}, {Padmanabhan},
  {O'Connell} et~al.}}]{hoagarwal2013xdqsoz}
\bibinfo{author}{\bibfnamefont{S.}~\bibnamefont{{Ho}}},
  \bibinfo{author}{\bibfnamefont{N.}~\bibnamefont{{Agarwal}}},
  \bibinfo{author}{\bibfnamefont{A.~D.} \bibnamefont{{Myers}}},
  \bibinfo{author}{\bibfnamefont{R.}~\bibnamefont{{Lyons}}},
  \bibinfo{author}{\bibfnamefont{A.}~\bibnamefont{{Disbrow}}},
  \bibinfo{author}{\bibfnamefont{H.-J.} \bibnamefont{{Seo}}},
  \bibinfo{author}{\bibfnamefont{A.}~\bibnamefont{{Ross}}},
  \bibinfo{author}{\bibfnamefont{C.}~\bibnamefont{{Hirata}}},
  \bibinfo{author}{\bibfnamefont{N.}~\bibnamefont{{Padmanabhan}}},
  \bibinfo{author}{\bibfnamefont{R.}~\bibnamefont{{O'Connell}}},
  \bibnamefont{et~al.}, \bibinfo{journal}{ArXiv e-prints}
  (\bibinfo{year}{2013}), \eprint{1311.2597}.

\bibitem[{\citenamefont{{Bovy} et~al.}(2011)\citenamefont{{Bovy}, {Hennawi},
  {Hogg}, {Myers}, {Kirkpatrick}, {Schlegel}, {Ross}, {Sheldon}, {McGreer},
  {Schneider} et~al.}}]{Bovy2010xdqso}
\bibinfo{author}{\bibfnamefont{J.}~\bibnamefont{{Bovy}}},
  \bibinfo{author}{\bibfnamefont{J.~F.} \bibnamefont{{Hennawi}}},
  \bibinfo{author}{\bibfnamefont{D.~W.} \bibnamefont{{Hogg}}},
  \bibinfo{author}{\bibfnamefont{A.~D.} \bibnamefont{{Myers}}},
  \bibinfo{author}{\bibfnamefont{J.~A.} \bibnamefont{{Kirkpatrick}}},
  \bibinfo{author}{\bibfnamefont{D.~J.} \bibnamefont{{Schlegel}}},
  \bibinfo{author}{\bibfnamefont{N.~P.} \bibnamefont{{Ross}}},
  \bibinfo{author}{\bibfnamefont{E.~S.} \bibnamefont{{Sheldon}}},
  \bibinfo{author}{\bibfnamefont{I.~D.} \bibnamefont{{McGreer}}},
  \bibinfo{author}{\bibfnamefont{D.~P.} \bibnamefont{{Schneider}}},
  \bibnamefont{et~al.}, \bibinfo{journal}{Astrophys.\ J.}
  \textbf{\bibinfo{volume}{729}}, \bibinfo{pages}{141} (\bibinfo{year}{2011}),
  \eprint{1011.6392}.

\bibitem[{\citenamefont{{Tegmark}}(1997)}]{Teg97}
\bibinfo{author}{\bibfnamefont{M.}~\bibnamefont{{Tegmark}}},
  \bibinfo{journal}{Phys.\ Rev.\ D.} \textbf{\bibinfo{volume}{55}},
  \bibinfo{pages}{5895} (\bibinfo{year}{1997}), \eprint{astro-ph/9611174}.

\bibitem[{\citenamefont{{Bond} et~al.}(1998)\citenamefont{{Bond}, {Jaffe}, and
  {Knox}}}]{BJK98b}
\bibinfo{author}{\bibfnamefont{J.~R.} \bibnamefont{{Bond}}},
  \bibinfo{author}{\bibfnamefont{A.~H.} \bibnamefont{{Jaffe}}},
  \bibnamefont{and} \bibinfo{author}{\bibfnamefont{L.}~\bibnamefont{{Knox}}},
  \bibinfo{journal}{Phys.\ Rev.\ D.} \textbf{\bibinfo{volume}{57}},
  \bibinfo{pages}{2117} (\bibinfo{year}{1998}), \eprint{astro-ph/9708203}.

\bibitem[{\citenamefont{{Tegmark} et~al.}(1998)\citenamefont{{Tegmark},
  {Hamilton}, {Strauss}, {Vogeley}, and {Szalay}}}]{THS1998future}
\bibinfo{author}{\bibfnamefont{M.}~\bibnamefont{{Tegmark}}},
  \bibinfo{author}{\bibfnamefont{A.~J.~S.} \bibnamefont{{Hamilton}}},
  \bibinfo{author}{\bibfnamefont{M.~A.} \bibnamefont{{Strauss}}},
  \bibinfo{author}{\bibfnamefont{M.~S.} \bibnamefont{{Vogeley}}},
  \bibnamefont{and} \bibinfo{author}{\bibfnamefont{A.~S.}
  \bibnamefont{{Szalay}}}, \bibinfo{journal}{Astrophys.\ J.}
  \textbf{\bibinfo{volume}{499}}, \bibinfo{pages}{555} (\bibinfo{year}{1998}),
  \eprint{astro-ph/9708020}.

\bibitem[{\citenamefont{Wandelt et~al.}(2004)\citenamefont{Wandelt, Larson, and
  Lakshminarayanan}}]{Wandelt:2003uk}
\bibinfo{author}{\bibfnamefont{B.~D.} \bibnamefont{Wandelt}},
  \bibinfo{author}{\bibfnamefont{D.~L.} \bibnamefont{Larson}},
  \bibnamefont{and}
  \bibinfo{author}{\bibfnamefont{A.}~\bibnamefont{Lakshminarayanan}},
  \bibinfo{journal}{Phys.Rev.} \textbf{\bibinfo{volume}{D70}},
  \bibinfo{pages}{083511} (\bibinfo{year}{2004}), \eprint{astro-ph/0310080}.

\bibitem[{\citenamefont{{Slosar} et~al.}(2004)\citenamefont{{Slosar}, {Seljak},
  and {Makarov}}}]{SlosarSeljak2004modeproj}
\bibinfo{author}{\bibfnamefont{A.}~\bibnamefont{{Slosar}}},
  \bibinfo{author}{\bibfnamefont{U.}~\bibnamefont{{Seljak}}}, \bibnamefont{and}
  \bibinfo{author}{\bibfnamefont{A.}~\bibnamefont{{Makarov}}},
  \bibinfo{journal}{Phys.\ Rev.\ D.} \textbf{\bibinfo{volume}{69}},
  \bibinfo{eid}{123003} (\bibinfo{year}{2004}), \eprint{astro-ph/0403073}.

\bibitem[{\citenamefont{{Wagner} and {Verde}}(2012)}]{WagnerVerde2012}
\bibinfo{author}{\bibfnamefont{C.}~\bibnamefont{{Wagner}}} \bibnamefont{and}
  \bibinfo{author}{\bibfnamefont{L.}~\bibnamefont{{Verde}}},
  \bibinfo{journal}{\jcap} \textbf{\bibinfo{volume}{3}}, \bibinfo{eid}{002}
  (\bibinfo{year}{2012}), \eprint{1102.3229}.

\bibitem[{\citenamefont{{Agullo} and {Shandera}}(2012)}]{2012JCAP...09..007A}
\bibinfo{author}{\bibfnamefont{I.}~\bibnamefont{{Agullo}}} \bibnamefont{and}
  \bibinfo{author}{\bibfnamefont{S.}~\bibnamefont{{Shandera}}},
  \bibinfo{journal}{\jcap} \textbf{\bibinfo{volume}{9}}, \bibinfo{eid}{007}
  (\bibinfo{year}{2012}), \eprint{1204.4409}.

\bibitem[{\citenamefont{{Dias} et~al.}(2013)\citenamefont{{Dias}, {Ribeiro},
  and {Seery}}}]{2013PhRvD..87j7301D}
\bibinfo{author}{\bibfnamefont{M.}~\bibnamefont{{Dias}}},
  \bibinfo{author}{\bibfnamefont{R.~H.} \bibnamefont{{Ribeiro}}},
  \bibnamefont{and} \bibinfo{author}{\bibfnamefont{D.}~\bibnamefont{{Seery}}},
  \bibinfo{journal}{\prd} \textbf{\bibinfo{volume}{87}}, \bibinfo{eid}{107301}
  (\bibinfo{year}{2013}), \eprint{1303.6000}.

\bibitem[{\citenamefont{{Agarwal} et~al.}(2013)\citenamefont{{Agarwal}, {Ho},
  and {Shandera}}}]{agarwalho2013xdqsoz}
\bibinfo{author}{\bibfnamefont{N.}~\bibnamefont{{Agarwal}}},
  \bibinfo{author}{\bibfnamefont{S.}~\bibnamefont{{Ho}}}, \bibnamefont{and}
  \bibinfo{author}{\bibfnamefont{S.}~\bibnamefont{{Shandera}}},
  \bibinfo{journal}{ArXiv e-prints}  (\bibinfo{year}{2013}),
  \eprint{1311.2606}.

\bibitem[{\citenamefont{Becker et~al.}(2011)\citenamefont{Becker, Huterer, and
  Kadota}}]{Becker:2010hx}
\bibinfo{author}{\bibfnamefont{A.}~\bibnamefont{Becker}},
  \bibinfo{author}{\bibfnamefont{D.}~\bibnamefont{Huterer}}, \bibnamefont{and}
  \bibinfo{author}{\bibfnamefont{K.}~\bibnamefont{Kadota}},
  \bibinfo{journal}{JCAP} \textbf{\bibinfo{volume}{1101}}, \bibinfo{pages}{006}
  (\bibinfo{year}{2011}), \eprint{1009.4189}.

\bibitem[{\citenamefont{Becker et~al.}(2012)\citenamefont{Becker, Huterer, and
  Kadota}}]{Becker:2012yr}
\bibinfo{author}{\bibfnamefont{A.}~\bibnamefont{Becker}},
  \bibinfo{author}{\bibfnamefont{D.}~\bibnamefont{Huterer}}, \bibnamefont{and}
  \bibinfo{author}{\bibfnamefont{K.}~\bibnamefont{Kadota}},
  \bibinfo{journal}{JCAP} \textbf{\bibinfo{volume}{1212}}, \bibinfo{pages}{034}
  (\bibinfo{year}{2012}), \eprint{1206.6165}.

\bibitem[{\citenamefont{{Myers} et~al.}(2006)\citenamefont{{Myers}, {Brunner},
  {Richards}, {Nichol}, {Schneider}, {Vanden Berk}, {Scranton}, {Gray}, and
  {Brinkmann}}}]{Myers2006first}
\bibinfo{author}{\bibfnamefont{A.~D.} \bibnamefont{{Myers}}},
  \bibinfo{author}{\bibfnamefont{R.~J.} \bibnamefont{{Brunner}}},
  \bibinfo{author}{\bibfnamefont{G.~T.} \bibnamefont{{Richards}}},
  \bibinfo{author}{\bibfnamefont{R.~C.} \bibnamefont{{Nichol}}},
  \bibinfo{author}{\bibfnamefont{D.~P.} \bibnamefont{{Schneider}}},
  \bibinfo{author}{\bibfnamefont{D.~E.} \bibnamefont{{Vanden Berk}}},
  \bibinfo{author}{\bibfnamefont{R.}~\bibnamefont{{Scranton}}},
  \bibinfo{author}{\bibfnamefont{A.~G.} \bibnamefont{{Gray}}},
  \bibnamefont{and}
  \bibinfo{author}{\bibfnamefont{J.}~\bibnamefont{{Brinkmann}}},
  \bibinfo{journal}{Astrophys.\ J.} \textbf{\bibinfo{volume}{638}},
  \bibinfo{pages}{622} (\bibinfo{year}{2006}), \eprint{arXiv:astro-ph/0510371}.

\bibitem[{\citenamefont{{Myers}
  et~al.}(2007{\natexlab{a}})\citenamefont{{Myers}, {Brunner}, {Nichol},
  {Richards}, {Schneider}, and {Bahcall}}}]{Myers2007one}
\bibinfo{author}{\bibfnamefont{A.~D.} \bibnamefont{{Myers}}},
  \bibinfo{author}{\bibfnamefont{R.~J.} \bibnamefont{{Brunner}}},
  \bibinfo{author}{\bibfnamefont{R.~C.} \bibnamefont{{Nichol}}},
  \bibinfo{author}{\bibfnamefont{G.~T.} \bibnamefont{{Richards}}},
  \bibinfo{author}{\bibfnamefont{D.~P.} \bibnamefont{{Schneider}}},
  \bibnamefont{and} \bibinfo{author}{\bibfnamefont{N.~A.}
  \bibnamefont{{Bahcall}}}, \bibinfo{journal}{Astrophys.\ J.}
  \textbf{\bibinfo{volume}{658}}, \bibinfo{pages}{85}
  (\bibinfo{year}{2007}{\natexlab{a}}), \eprint{arXiv:astro-ph/0612190}.

\bibitem[{\citenamefont{{Myers}
  et~al.}(2007{\natexlab{b}})\citenamefont{{Myers}, {Brunner}, {Richards},
  {Nichol}, {Schneider}, and {Bahcall}}}]{Myers2007two}
\bibinfo{author}{\bibfnamefont{A.~D.} \bibnamefont{{Myers}}},
  \bibinfo{author}{\bibfnamefont{R.~J.} \bibnamefont{{Brunner}}},
  \bibinfo{author}{\bibfnamefont{G.~T.} \bibnamefont{{Richards}}},
  \bibinfo{author}{\bibfnamefont{R.~C.} \bibnamefont{{Nichol}}},
  \bibinfo{author}{\bibfnamefont{D.~P.} \bibnamefont{{Schneider}}},
  \bibnamefont{and} \bibinfo{author}{\bibfnamefont{N.~A.}
  \bibnamefont{{Bahcall}}}, \bibinfo{journal}{Astrophys.\ J.}
  \textbf{\bibinfo{volume}{658}}, \bibinfo{pages}{99}
  (\bibinfo{year}{2007}{\natexlab{b}}), \eprint{arXiv:astro-ph/0612191}.

\bibitem[{\citenamefont{{Shen} et~al.}(2007)\citenamefont{{Shen}, {Strauss},
  {Oguri}, {Hennawi}, {Fan}, {Richards}, {Hall}, {Gunn}, {Schneider}, {Szalay}
  et~al.}}]{2007AJ....133.2222S}
\bibinfo{author}{\bibfnamefont{Y.}~\bibnamefont{{Shen}}},
  \bibinfo{author}{\bibfnamefont{M.~A.} \bibnamefont{{Strauss}}},
  \bibinfo{author}{\bibfnamefont{M.}~\bibnamefont{{Oguri}}},
  \bibinfo{author}{\bibfnamefont{J.~F.} \bibnamefont{{Hennawi}}},
  \bibinfo{author}{\bibfnamefont{X.}~\bibnamefont{{Fan}}},
  \bibinfo{author}{\bibfnamefont{G.~T.} \bibnamefont{{Richards}}},
  \bibinfo{author}{\bibfnamefont{P.~B.} \bibnamefont{{Hall}}},
  \bibinfo{author}{\bibfnamefont{J.~E.} \bibnamefont{{Gunn}}},
  \bibinfo{author}{\bibfnamefont{D.~P.} \bibnamefont{{Schneider}}},
  \bibinfo{author}{\bibfnamefont{A.~S.} \bibnamefont{{Szalay}}},
  \bibnamefont{et~al.}, \bibinfo{journal}{\apj} \textbf{\bibinfo{volume}{133}},
  \bibinfo{pages}{2222} (\bibinfo{year}{2007}), \eprint{astro-ph/0702214}.

\bibitem[{\citenamefont{{White} et~al.}(2012)\citenamefont{{White}, {Myers},
  {Ross}, {Schlegel}, {Hennawi}, {Shen}, {McGreer}, {Strauss}, {Bolton}, {Bovy}
  et~al.}}]{White2012specqso}
\bibinfo{author}{\bibfnamefont{M.}~\bibnamefont{{White}}},
  \bibinfo{author}{\bibfnamefont{A.~D.} \bibnamefont{{Myers}}},
  \bibinfo{author}{\bibfnamefont{N.~P.} \bibnamefont{{Ross}}},
  \bibinfo{author}{\bibfnamefont{D.~J.} \bibnamefont{{Schlegel}}},
  \bibinfo{author}{\bibfnamefont{J.~F.} \bibnamefont{{Hennawi}}},
  \bibinfo{author}{\bibfnamefont{Y.}~\bibnamefont{{Shen}}},
  \bibinfo{author}{\bibfnamefont{I.}~\bibnamefont{{McGreer}}},
  \bibinfo{author}{\bibfnamefont{M.~A.} \bibnamefont{{Strauss}}},
  \bibinfo{author}{\bibfnamefont{A.~S.} \bibnamefont{{Bolton}}},
  \bibinfo{author}{\bibfnamefont{J.}~\bibnamefont{{Bovy}}},
  \bibnamefont{et~al.}, \bibinfo{journal}{Mon.\ Not.\ Roy.\ Astron.\ Soc.}
  \textbf{\bibinfo{volume}{424}}, \bibinfo{pages}{933} (\bibinfo{year}{2012}),
  \eprint{1203.5306}.

\bibitem[{\citenamefont{{Porciani} and {Norberg}}(2006)}]{porcianiNorberg2006}
\bibinfo{author}{\bibfnamefont{C.}~\bibnamefont{{Porciani}}} \bibnamefont{and}
  \bibinfo{author}{\bibfnamefont{P.}~\bibnamefont{{Norberg}}},
  \bibinfo{journal}{Mon.\ Not.\ Roy.\ Astron.\ Soc.}
  \textbf{\bibinfo{volume}{371}}, \bibinfo{pages}{1824} (\bibinfo{year}{2006}),
  \eprint{astro-ph/0607348}.

\bibitem[{\citenamefont{{Croom} et~al.}(2005)\citenamefont{{Croom}, {Boyle},
  {Shanks}, {Smith}, {Miller}, {Outram}, {Loaring}, {Hoyle}, and {da
  {\^A}ngela}}}]{2005MNRAS.356..415C}
\bibinfo{author}{\bibfnamefont{S.~M.} \bibnamefont{{Croom}}},
  \bibinfo{author}{\bibfnamefont{B.~J.} \bibnamefont{{Boyle}}},
  \bibinfo{author}{\bibfnamefont{T.}~\bibnamefont{{Shanks}}},
  \bibinfo{author}{\bibfnamefont{R.~J.} \bibnamefont{{Smith}}},
  \bibinfo{author}{\bibfnamefont{L.}~\bibnamefont{{Miller}}},
  \bibinfo{author}{\bibfnamefont{P.~J.} \bibnamefont{{Outram}}},
  \bibinfo{author}{\bibfnamefont{N.~S.} \bibnamefont{{Loaring}}},
  \bibinfo{author}{\bibfnamefont{F.}~\bibnamefont{{Hoyle}}}, \bibnamefont{and}
  \bibinfo{author}{\bibfnamefont{J.}~\bibnamefont{{da {\^A}ngela}}},
  \bibinfo{journal}{\mnras} \textbf{\bibinfo{volume}{356}},
  \bibinfo{pages}{415} (\bibinfo{year}{2005}), \eprint{astro-ph/0409314}.

\bibitem[{\citenamefont{{Ross} et~al.}(2009)\citenamefont{{Ross}, {Shen},
  {Strauss}, {Vanden Berk}, {Connolly}, {Richards}, {Schneider}, {Weinberg},
  {Hall}, {Bahcall} et~al.}}]{Ross2009specqsodr5}
\bibinfo{author}{\bibfnamefont{N.~P.} \bibnamefont{{Ross}}},
  \bibinfo{author}{\bibfnamefont{Y.}~\bibnamefont{{Shen}}},
  \bibinfo{author}{\bibfnamefont{M.~A.} \bibnamefont{{Strauss}}},
  \bibinfo{author}{\bibfnamefont{D.~E.} \bibnamefont{{Vanden Berk}}},
  \bibinfo{author}{\bibfnamefont{A.~J.} \bibnamefont{{Connolly}}},
  \bibinfo{author}{\bibfnamefont{G.~T.} \bibnamefont{{Richards}}},
  \bibinfo{author}{\bibfnamefont{D.~P.} \bibnamefont{{Schneider}}},
  \bibinfo{author}{\bibfnamefont{D.~H.} \bibnamefont{{Weinberg}}},
  \bibinfo{author}{\bibfnamefont{P.~B.} \bibnamefont{{Hall}}},
  \bibinfo{author}{\bibfnamefont{N.~A.} \bibnamefont{{Bahcall}}},
  \bibnamefont{et~al.}, \bibinfo{journal}{Astrophys.\ J.}
  \textbf{\bibinfo{volume}{697}}, \bibinfo{pages}{1634} (\bibinfo{year}{2009}),
  \eprint{0903.3230}.

\bibitem[{\citenamefont{{Challinor} and
  {Lewis}}(2011)}]{challinorlewis2011cambsources}
\bibinfo{author}{\bibfnamefont{A.}~\bibnamefont{{Challinor}}} \bibnamefont{and}
  \bibinfo{author}{\bibfnamefont{A.}~\bibnamefont{{Lewis}}},
  \bibinfo{journal}{Phys.\ Rev.\ D.} \textbf{\bibinfo{volume}{84}},
  \bibinfo{eid}{043516} (\bibinfo{year}{2011}), \eprint{1105.5292}.

\bibitem[{\citenamefont{{Foreman-Mackey}
  et~al.}(2013)\citenamefont{{Foreman-Mackey}, {Hogg}, {Lang}, and
  {Goodman}}}]{ForemanMackey2013emcee}
\bibinfo{author}{\bibfnamefont{D.}~\bibnamefont{{Foreman-Mackey}}},
  \bibinfo{author}{\bibfnamefont{D.~W.} \bibnamefont{{Hogg}}},
  \bibinfo{author}{\bibfnamefont{D.}~\bibnamefont{{Lang}}}, \bibnamefont{and}
  \bibinfo{author}{\bibfnamefont{J.}~\bibnamefont{{Goodman}}},
  \bibinfo{journal}{Publications of the Astronomical Society of the Pacific}
  \textbf{\bibinfo{volume}{125}}, \bibinfo{pages}{306} (\bibinfo{year}{2013}),
  \eprint{1202.3665}.

\bibitem[{\citenamefont{{Planck
  Collaboration}}(2013{\natexlab{b}})}]{Planck2013cosmologicalparams}
\bibinfo{author}{\bibnamefont{{Planck Collaboration}}}, \bibinfo{journal}{ArXiv
  e-prints}  (\bibinfo{year}{2013}{\natexlab{b}}), \eprint{1303.5076}.

\bibitem[{\citenamefont{Roth and Porciani}(2012)}]{Roth:2012yy}
\bibinfo{author}{\bibfnamefont{N.}~\bibnamefont{Roth}} \bibnamefont{and}
  \bibinfo{author}{\bibfnamefont{C.}~\bibnamefont{Porciani}},
  \bibinfo{journal}{Mon.Not.Roy.Astron.Soc.} \textbf{\bibinfo{volume}{425}},
  \bibinfo{pages}{L81} (\bibinfo{year}{2012}), \eprint{1205.3165}.

\bibitem[{\citenamefont{{LSST Dark Energy Science
  Collaboration}}(2012)}]{2012arXiv1211.0310L}
\bibinfo{author}{\bibnamefont{{LSST Dark Energy Science Collaboration}}},
  \bibinfo{journal}{ArXiv e-prints}  (\bibinfo{year}{2012}),
  \eprint{1211.0310}.

\bibitem[{\citenamefont{{G{\'o}rski} et~al.}(2005)\citenamefont{{G{\'o}rski},
  {Hivon}, {Banday}, {Wandelt}, {Hansen}, {Reinecke}, and
  {Bartelmann}}}]{healpix1}
\bibinfo{author}{\bibfnamefont{K.~M.} \bibnamefont{{G{\'o}rski}}},
  \bibinfo{author}{\bibfnamefont{E.}~\bibnamefont{{Hivon}}},
  \bibinfo{author}{\bibfnamefont{A.~J.} \bibnamefont{{Banday}}},
  \bibinfo{author}{\bibfnamefont{B.~D.} \bibnamefont{{Wandelt}}},
  \bibinfo{author}{\bibfnamefont{F.~K.} \bibnamefont{{Hansen}}},
  \bibinfo{author}{\bibfnamefont{M.}~\bibnamefont{{Reinecke}}},
  \bibnamefont{and}
  \bibinfo{author}{\bibfnamefont{M.}~\bibnamefont{{Bartelmann}}},
  \bibinfo{journal}{Astrophys.\ J.} \textbf{\bibinfo{volume}{622}},
  \bibinfo{pages}{759} (\bibinfo{year}{2005}), \eprint{astro-ph/0409513}.

\end{thebibliography}

\end{document}